\documentclass{PoS}

\graphicspath{{figures/}}

\title{$SU(3)$ quark-antiquark QCD flux tube}

\ShortTitle{$SU(3)$ quark-antiquark QCD flux tube}

\author{\speaker{Pedro Bicudo} \\  
        CFTP, Departamento de F\'{\i}sica, Instituto Superior T\'ecnico,
        Universidade T\'ecnica de Lisboa \\
        E-mail: \email{bicudo@ist.utl.pt}
}

\author{Marco Cardoso\\
       CFTP, Departamento de F\'{\i}sica, Instituto Superior T\'ecnico,
       Universidade T\'ecnica de Lisboa \\
       E-mail: \email{mjdcc@cftp.ist.utl.pt}
}

\author{Nuno Cardoso\\
       CFTP, Departamento de F\'{\i}sica, Instituto Superior T\'ecnico,
       Universidade T\'ecnica de Lisboa \\
       E-mail: \email{nunocardoso@cftp.ist.utl.pt}
}

\abstract{
We compute the quark-antiquark flux tube for pure gauge SU(3) in space-time 3 + 1 dimensions.
To increase the signal over noise ratio, we apply the improved multihit and extended smearing techniques.
We fit the field densities with an appropriate ansatz and we observe both the screening of the color fields
and the quantum widening of the flux tube in the mediator plane and in the charge planes.
}

\FullConference{31st International Symposium on Lattice Field Theory - LATTICE 2013\\
		July 29 - August 3, 2013\\
		Mainz, Germany}

\begin{document}

\section{Introduction}

Confinement is a central feature of strong interactions. One of it's aspects is the formation of a color flux-tube 
between a quark and an antiquark in a meson. Here we study the profile of a mesonic flux tube.

Quantum string models based on the Nambu-Goto \cite{Nambu:1978bd,Goto:1971ce}
action $S = - \sigma \int d^2\Sigma$, predict a Gaussian profile of the flux-tube,
with a logarithmic increase \cite{Luscher:1980iy} of the the squared width of the quark-antiquark flux-tube
\begin{equation}
	w^2 \sim w_0^2 \log( \frac{R}{R_0} )
\end{equation}
while from models based on superconductivity \cite{Nambu:1974zg} it is naturally expected an exponential decay of the flux-tube,
with the length parameter $\lambda$ akin to the London penetration length, inverse of a dual gluon mass $\mu$.

\section{Computation of the Chromo-fields}

We calculate the chromo-fields with the correlation of the plaquette
\begin{equation}
P_{\mu\nu} = 1 - \frac{1}{3} \mbox{Tr}[U_\mu(s) U_\nu(s+\mu) U_\mu^\dagger(s+\nu) U_\nu^\dagger(s)]
\end{equation}
with the mesonic Wilson Loop operator $W$.

The fields are the given by
\begin{equation}
	\langle {B_i}^2(\mathbf r) \rangle	 = \frac{\langle W(R,T)\,P(\mathbf r)_{jk} \rangle}{\langle W(R,T) \rangle}
	- \langle P(\mathbf r)_{jk} \rangle
\end{equation}
\begin{equation}
	\langle {E_i}^2(\mathbf r) \rangle = \langle P(\mathbf r)_{0i} \rangle
	-\frac{\langle W(R,T) \,P(\mathbf r)_{0i}\rangle} {\langle{W(R,T)}\rangle}
\end{equation}
with the Lagrangian density being $\mathcal{L} = \frac{1}{2}(\mathbf{E}^2 - \mathbf{B}^2)$.

\section{Noise Reduction}

We need to improve the signal to noise ratio, in order to go to relatively large distances and obtain a sufficiently
clear results. To do so, we use several techniques.

\subsection{Extended Multihit}

Since the multilevel technique \cite{Luscher:2001up} is  very demanding computationally  and the multihit technique \cite{Parisi:1983hm}
 doesn't reduce sufficiently the errors, we utilize a different technique the extended multihit.
Here, instead of taking the thermal average of a temporal link with the first neighbors,
we fix the higher order neighbors, and apply the heat-bath algorithm to all the links inside, averaging the central link.
\begin{equation}
U_{4}\rightarrow\overline{U}_{4}=\frac{\int[\mathcal{D}U]_{\Omega}\, U_{4}\, e^{\beta\sum_{\mu\mathbf{s}}\mbox{Tr}[U_{\mu}(\mathbf{s})F_{\mu}^{\dagger}(\mathbf{\mathbf{s}})]}}{\int[\mathcal{D}U]_{\Omega}\, e^{\beta\sum_{\mu\mathbf{s}}\mbox{Tr}[U_{\mu}(\mathbf{s})F_{\mu}^{\dagger}]}}
\end{equation}


\begin{figure}
	\centering
     \includegraphics[bb=130bp 100bp 280bp 280bp,clip,width=0.20\columnwidth]{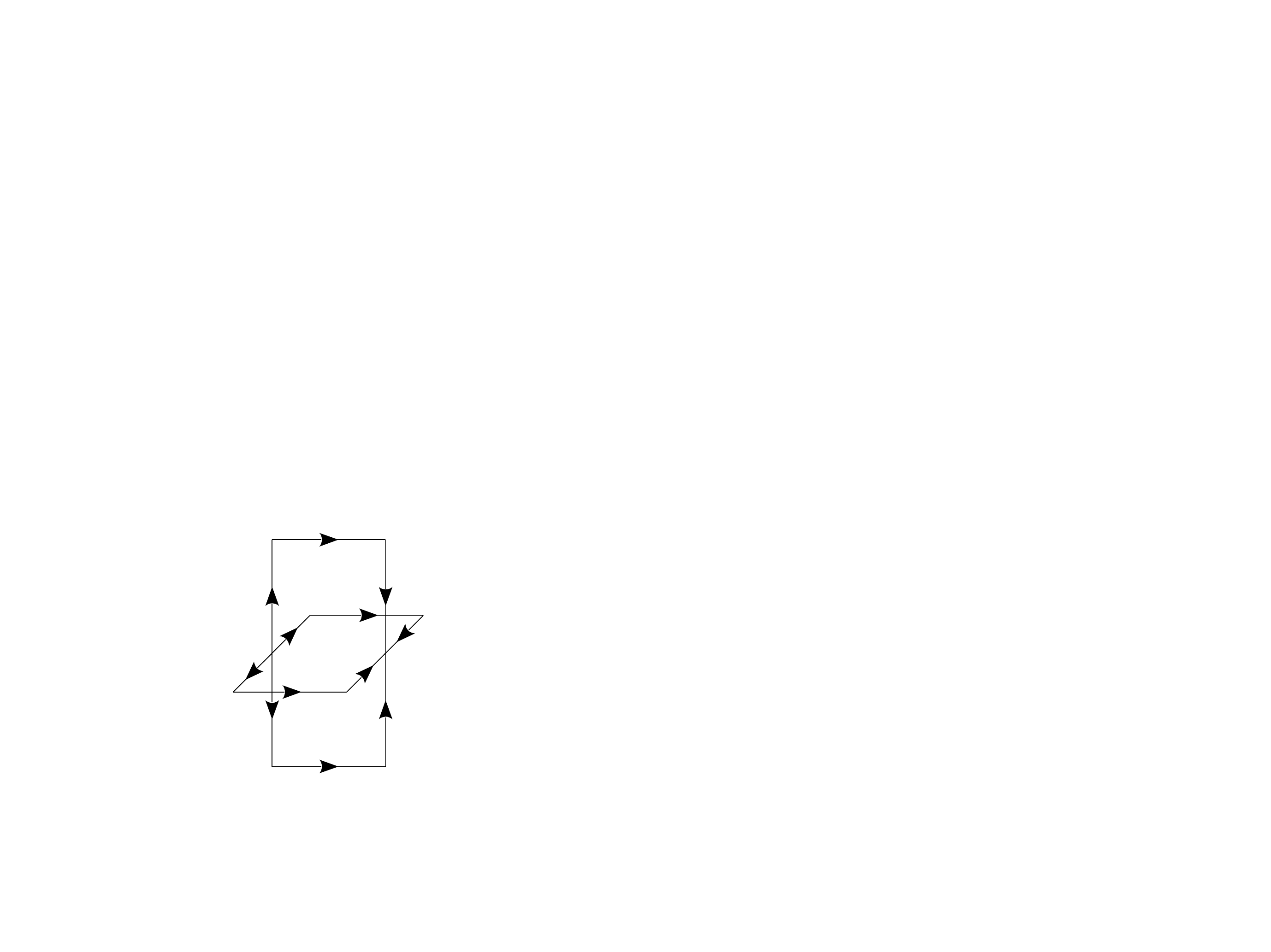}
     \includegraphics[bb=320bp 80bp 480bp 300bp,clip,width=0.22\columnwidth]{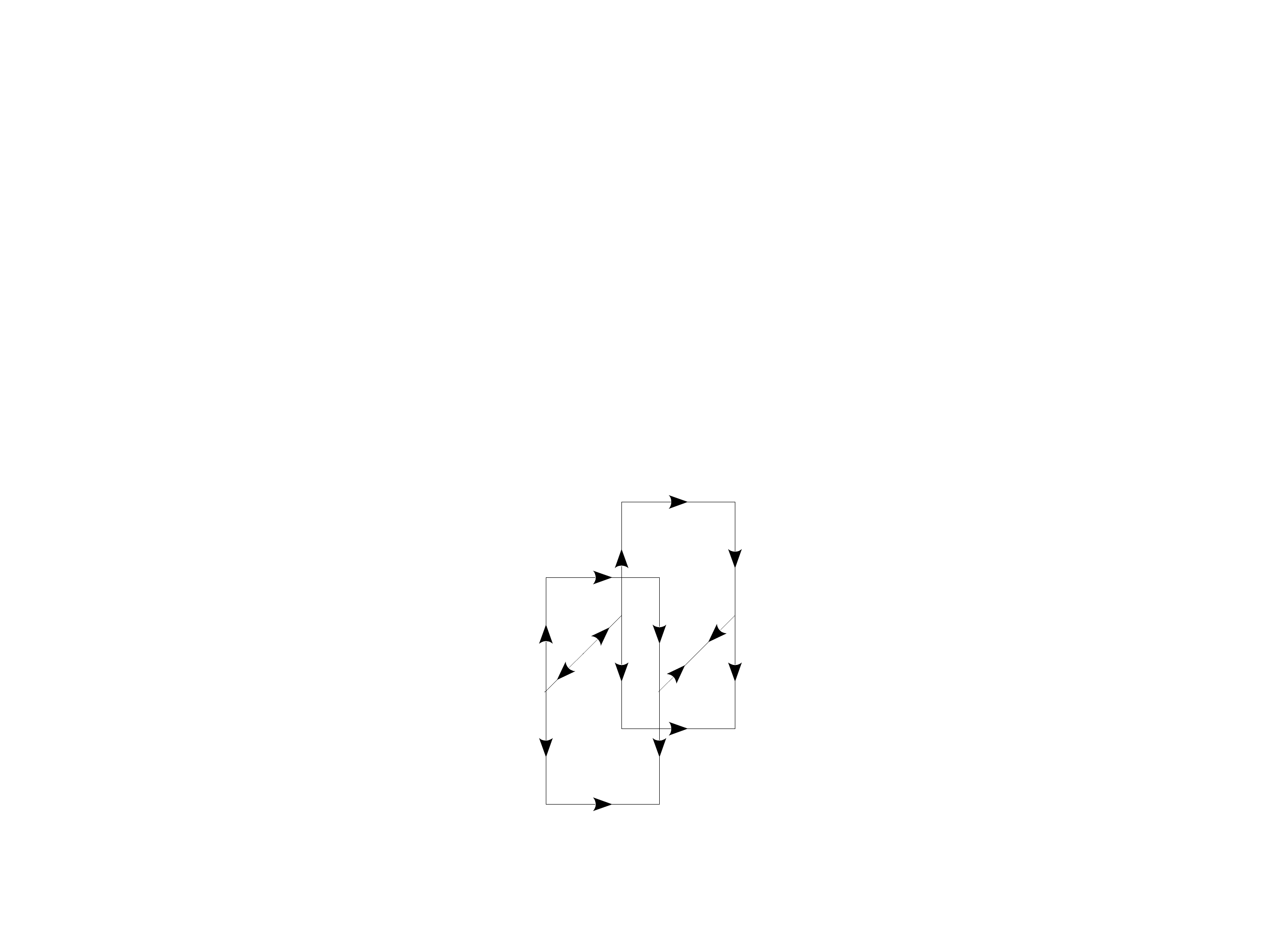}
     \includegraphics[bb=510bp 30bp 700bp 340bp,clip,width=0.26\columnwidth]{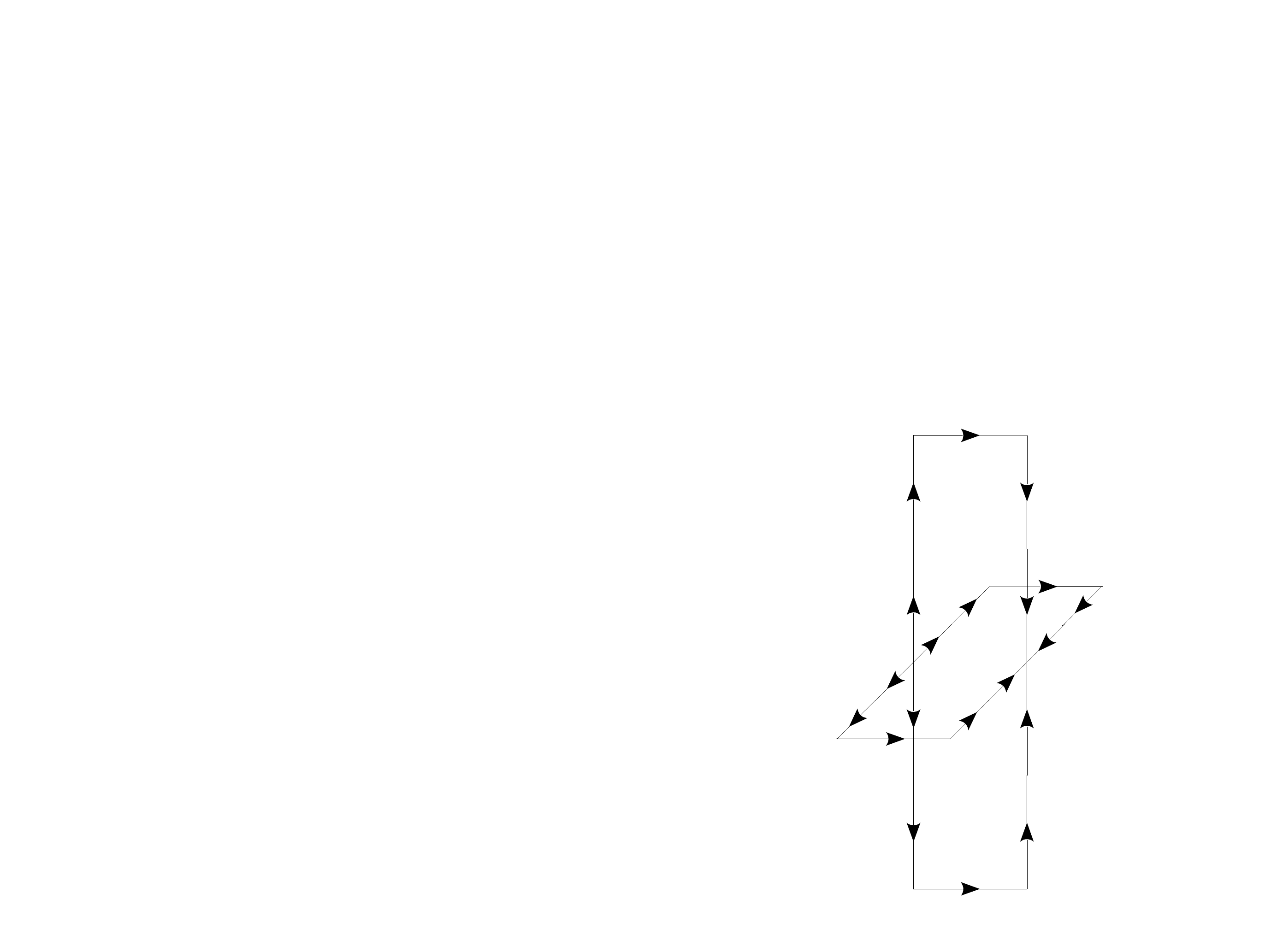}
	\caption{Staples used in the improved smearing. \label{staples}}
\end{figure}

\begin{figure}
	\centering
     \includegraphics[width=0.50\textwidth]{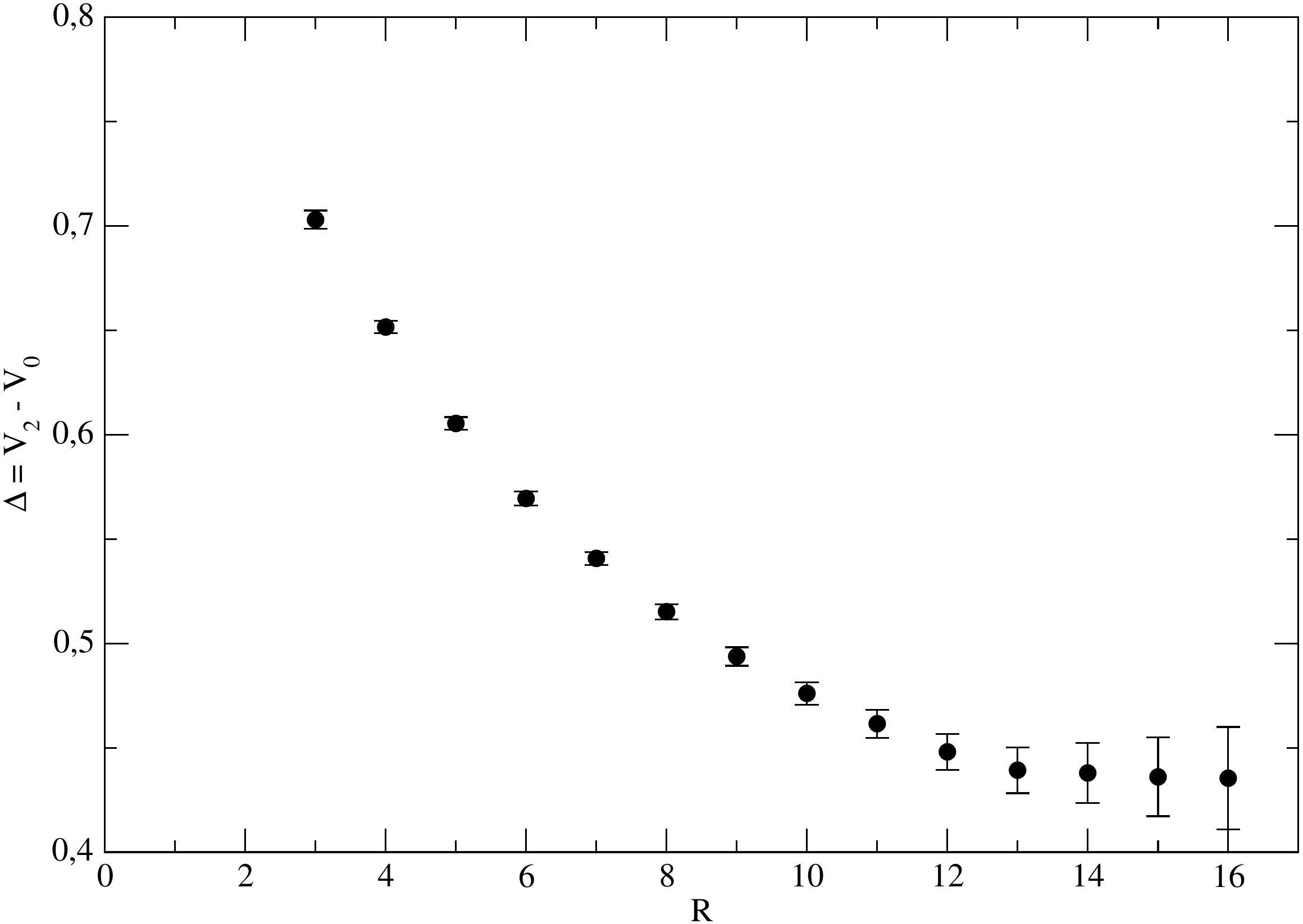}
	\caption{Energy gap as a function of quark-antiquark distance. \label{Delta}}
\end{figure}

\subsection{Extended Spatial Smearing}

Insted of using APE smearing we use, in order to further reduce the excited states contribution,
an improved smearing algorithm with higher order staples.
In this algorithm, each spatial link in the Wilson Loop, is replaced by (see Fig. \ref{staples})
\begin{equation}
U_{i}\rightarrow\mathcal{P}_{SU(3)}\Big[U_{i}+w_{1}\sum_{j}S_{ij}^{1}+w_{2}\sum_{j}S_{ij}^{2}+w_{3}\sum_{j}S_{ij}^{3}\Big]
\end{equation}
The plaquette is neither affected by this algorithm, or by Extended Multihit.

\subsection{Compute $\Delta$}

In order to minimize systematic errors arising from the small temporal extent of Wilson loops,
we calculate the fields by fitting the results with the formula
\begin{equation}
\langle F\rangle_{t}=\langle F\rangle_{\infty}+b\, e^{-\Delta t}
\end{equation}
where $\Delta$ is the energy gap between the ground state and the first excited state.
To calculate the gap we use a variational basis with four smearing states.
The results for the energy gap are shown in Fig. \ref{Delta}.

\section{Results}

\begin{figure}
\centering
\includegraphics[width=0.50\columnwidth]{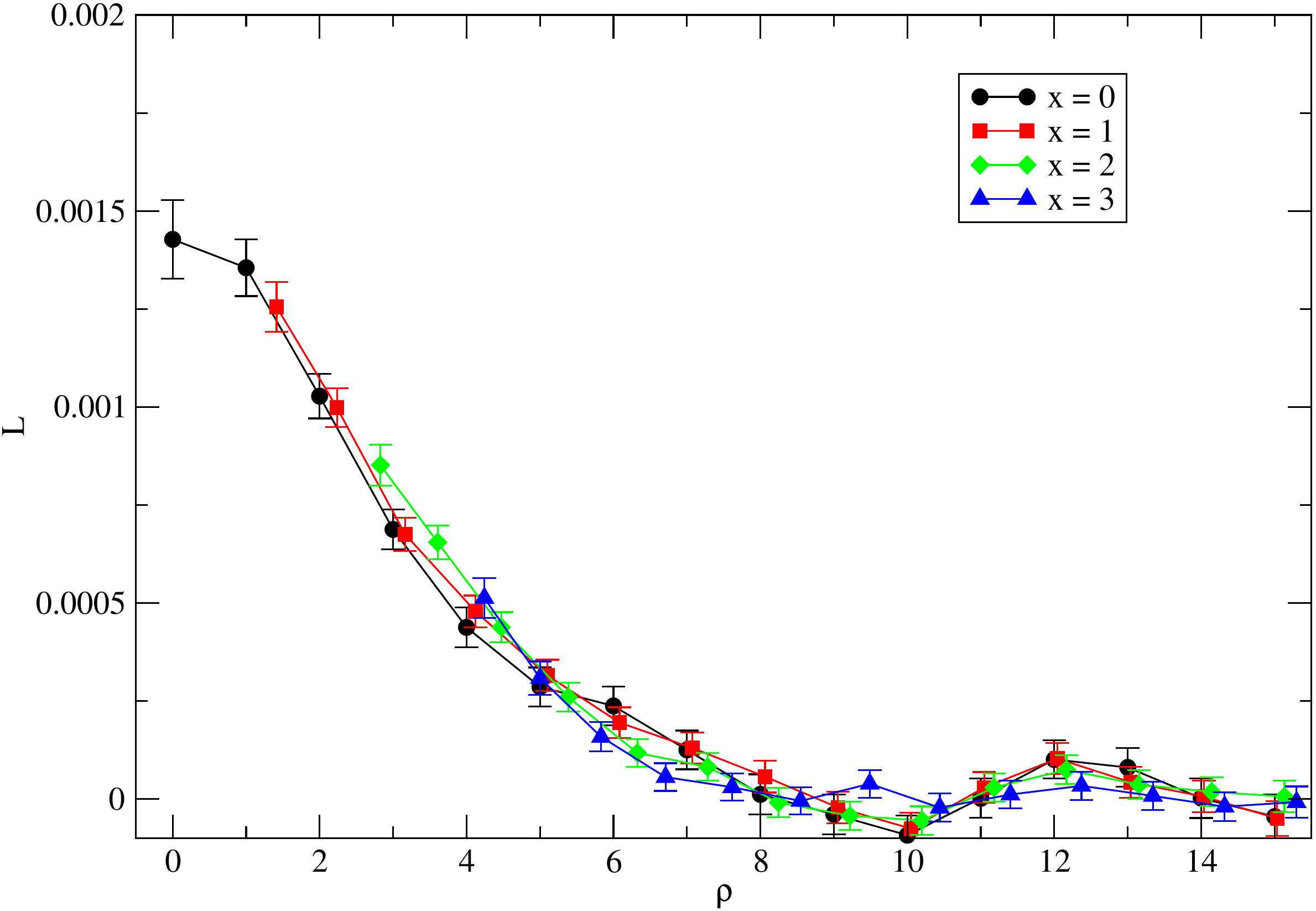}
\caption{Lagrangian density in the mediator plane. \label{Lx} }
\end{figure}

Results for the Lagrangian density in the quark-antiquark mediator plane are given in Fig. \ref{Lx} and,
as can be seen, with the noise reduction techniques we use, statistical errors are already smaller
than systematic errors that break rotational invariance.

In addition to calculating the fields in the mediator plane, we also compute them in the charge planes, defined as the planes that contains
the quark or the antiquark and is perpendicular to the quark-antiquark axis.
 
\begin{figure}
\centering
\includegraphics[width=0.60\columnwidth]{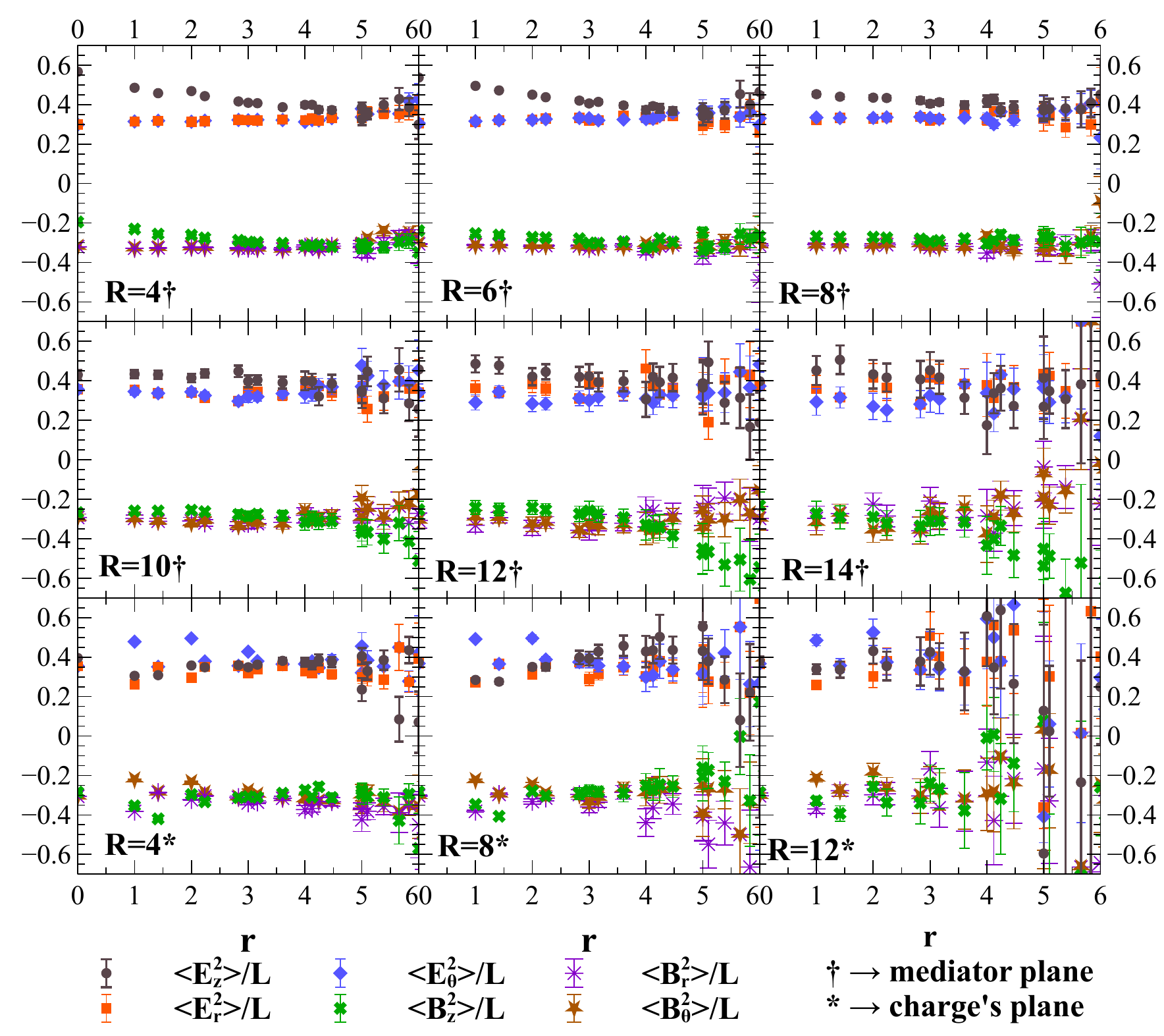}
\caption{Ratio of the squared field components to the Lagrangian density.} \label{ratios}
\end{figure}

In Fig. \ref{ratios} we show the ratio of the different field components to the Lagrangian density.
As can be seen the ratios are of the same order of magnitude, with the ${E_z}^2$ component begin larger
at small distances in the mediator plane, and the ${E_\theta}^2$ component being the more important
in the plane that contain the sources, close to them.
For sufficiently large distances the ratios are $\sim 0.4$ for the chromoelectric field components, and
$\sim 0.3$ for the chromomagnetic field components.
This means that, at sufficiently large distances the behaviors of the fields are essentially the same.

\begin{figure}
 \centering
 \includegraphics[width=0.60\columnwidth]{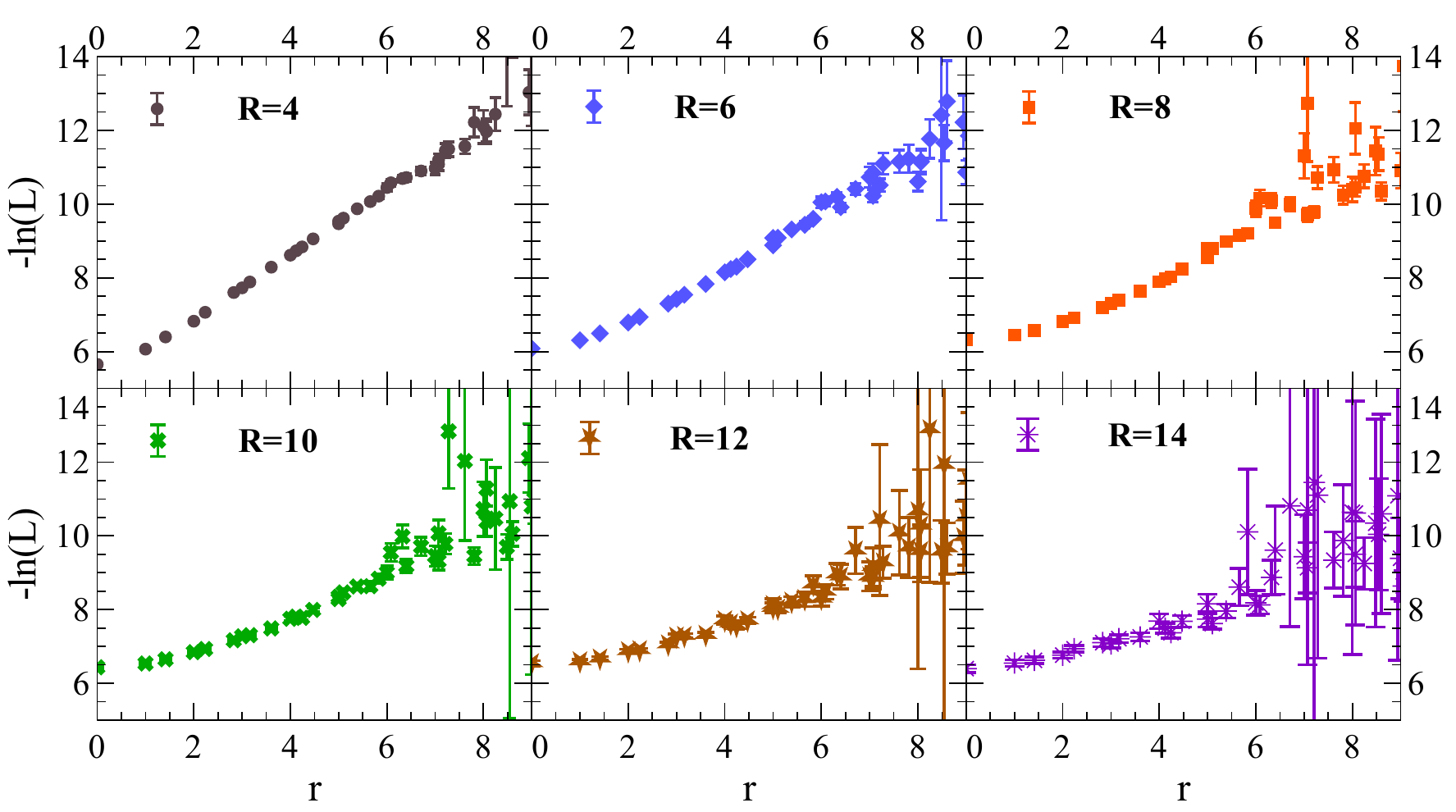}
 \caption{Logarithm of the Lagrangian density. } \label{logfig}
\end{figure}

Another interesting result is shown in Fig. \ref{logfig}. There, the logarithm of the Lagrangian density is
plotted against the distance to the center of the flux tube for the mediator plane.
As can be seen, the flux-tube has a Gaussian behavior close to the center but is exponential at large distances.
    
\subsection{Lagrangian density profiles}

\begin{figure}
	\centering
    \includegraphics[width=0.60\columnwidth]{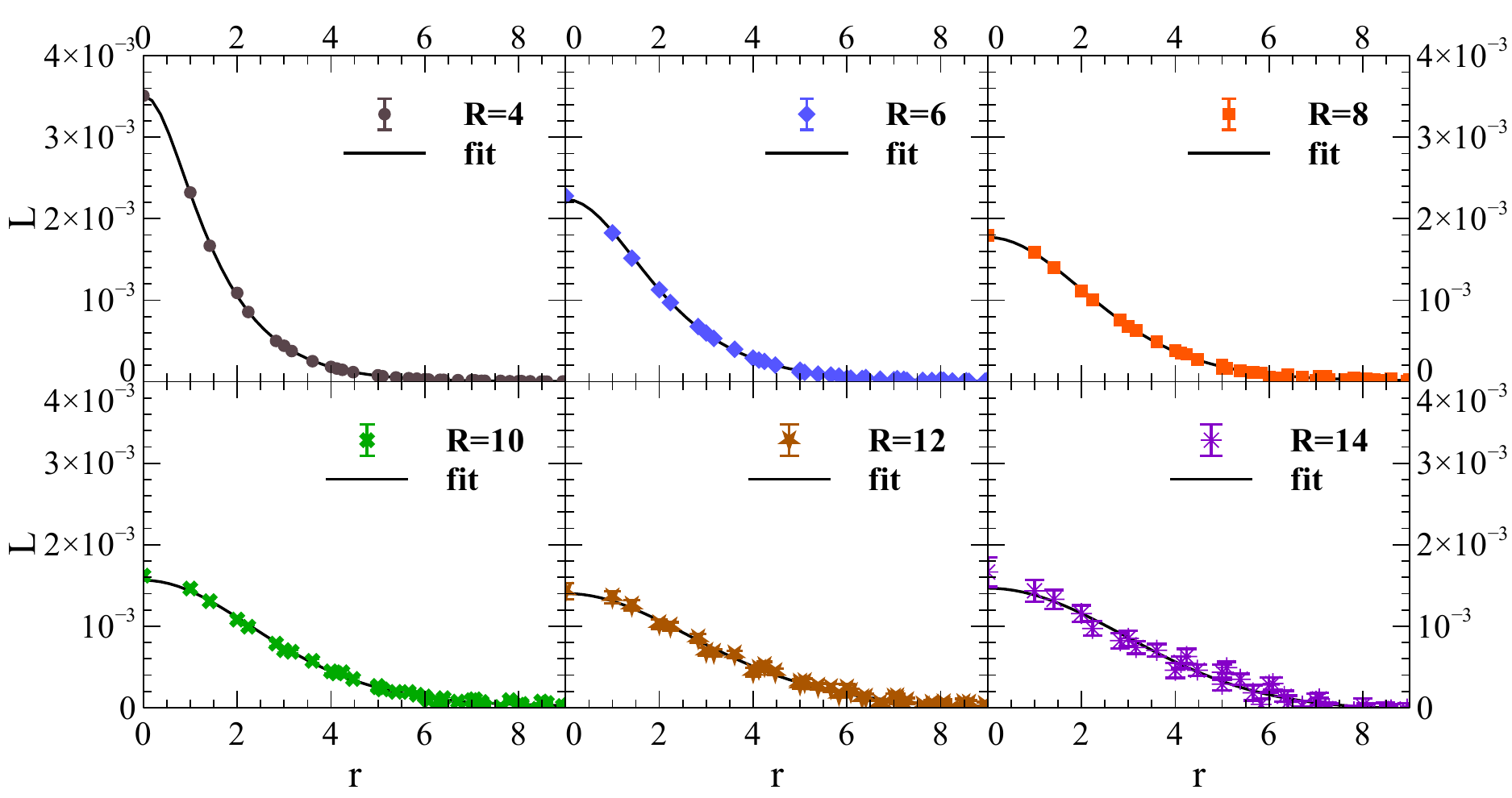}
	\caption{Lagrangian density profiles and fitted function.} \label{lagfits}
\end{figure}

In order to fit both the small and large distance part of the flux tube profile, we fit the data points for the
Lagrangian density shown in Fig. \ref{lagfits} to the ansatz
$\mathcal{L} = \mathcal{L}_0 \exp \left(-{2 \over \lambda} \sqrt{r^2 + \nu^2} +  2 {\nu \over \lambda} \right)$.
The obtained parameters are shown in Table \ref{fitres}.
     
\begin{table}
  \centering
  \begin{tabular}{c|cccc}
  $R \ [a]$ & $10^3 {\cal L}_0 $ & $\lambda \ [a]$ & $\nu \  [a]$ & $ \chi ^2 / dof$ \\
  \hline
  4 & 3.509 $\pm$ 26.72 & 2.165 $\pm$ 0.033 & 0.877 $\pm$ 3.335 & 4.086 \\
  6 & 2.236 $\pm$ 0.078 & 2.379 $\pm$ 0.156 & 2.04 $\pm$ 0.365 & 2.254 \\
  8 & 1.762 $\pm$ 0.023 & 2.052 $\pm$ 0.201 & 4.092 $\pm$ 20.22 & 1.999 \\
  10 & 1.549 $\pm$ 0.046 & 2.088 $\pm$ 0.536 & 5.306 $\pm$ 36.43 & 1.477 \\
  12 & 1.357 $\pm$ 0.051 & 0.913 $\pm$ 2.044 & 17.41 $\pm$ 200.1 & 1.055 \\
  14 & 1.491 $\pm$ 0.053 & 0.064 $\pm$ 0.018 & 268.0 $\pm$ 1392.4 & 1.331
  \end{tabular}
  \caption{Results for the fit parameters $\lambda$ and $\nu$. \label{fitres} }
\end{table}

\begin{figure}
\centering
\includegraphics[width=0.50\columnwidth]{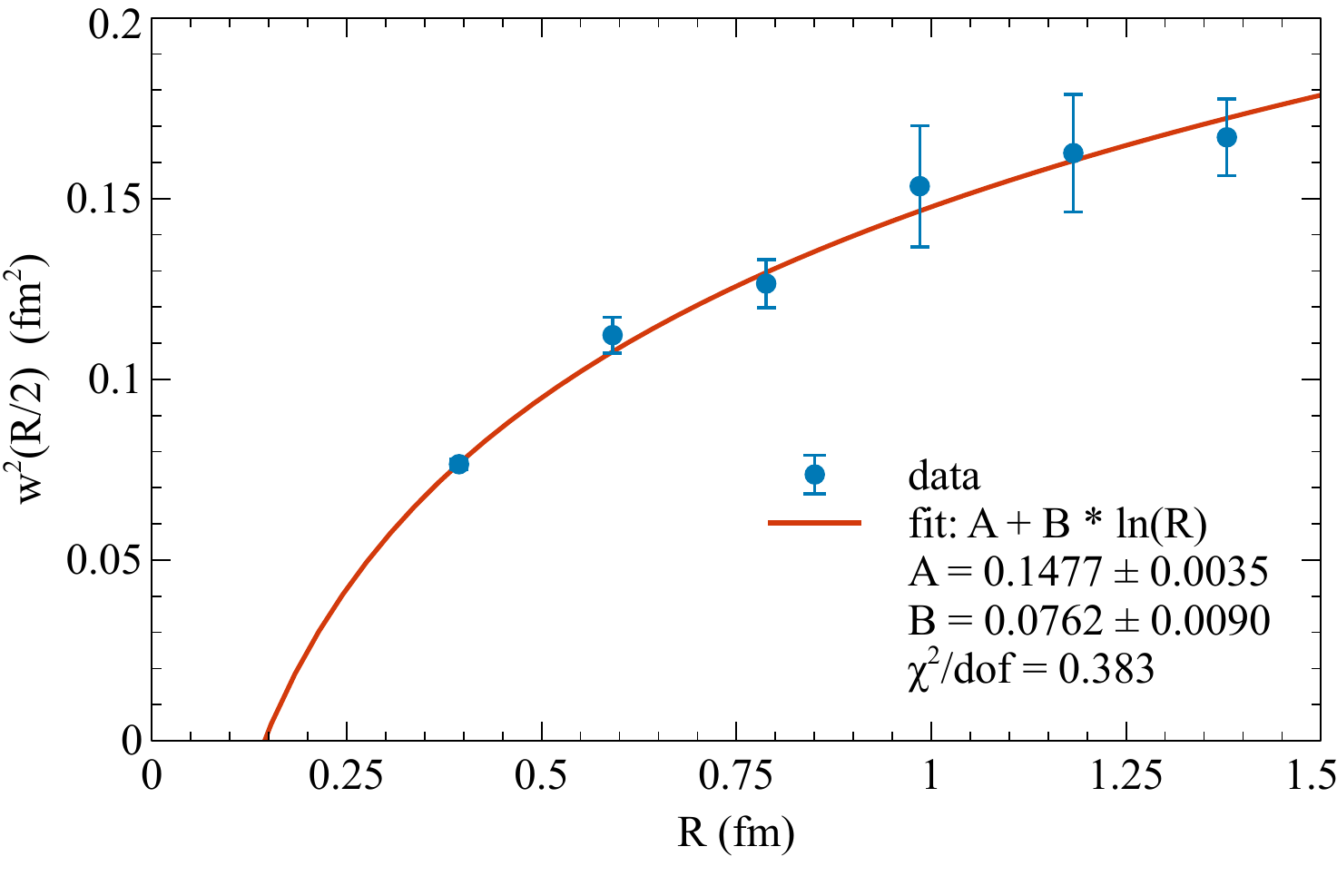}
\caption{Computed squared widths as a function of distance, and fit to logarithmic ansatz.} \label{log fit}
\end{figure}

Using this fit results, we directly calculate the width of the flux tube, analytically,
\begin{equation}
	 {\sqrt{ \langle r^2\rangle}} = 
\sqrt{ {3 \over 2} \lambda^2 + 2 {  \lambda  \nu^2  \over \lambda + 2  \nu } }  \ .
\label{width}
\end{equation}
the results for the squared widths, obtained from the fitted parameters, are shown in Table \ref{sqwidths}.
Then we try to fit the obtained squared widths with a logarithmic ansatz,  
\begin{equation}
	{\sqrt{ \langle r^2\rangle}} = A + B \log R \ . 
\end{equation}
Fitting these results, we find, as shown in Fig. \ref{log fit} that the squared width of the flux tube has a logarithmic increase as a function of the
quark-antiquark distance, in accordance with the widening hypothesis.

\begin{figure}
\centering
\includegraphics[width=0.60\textwidth]{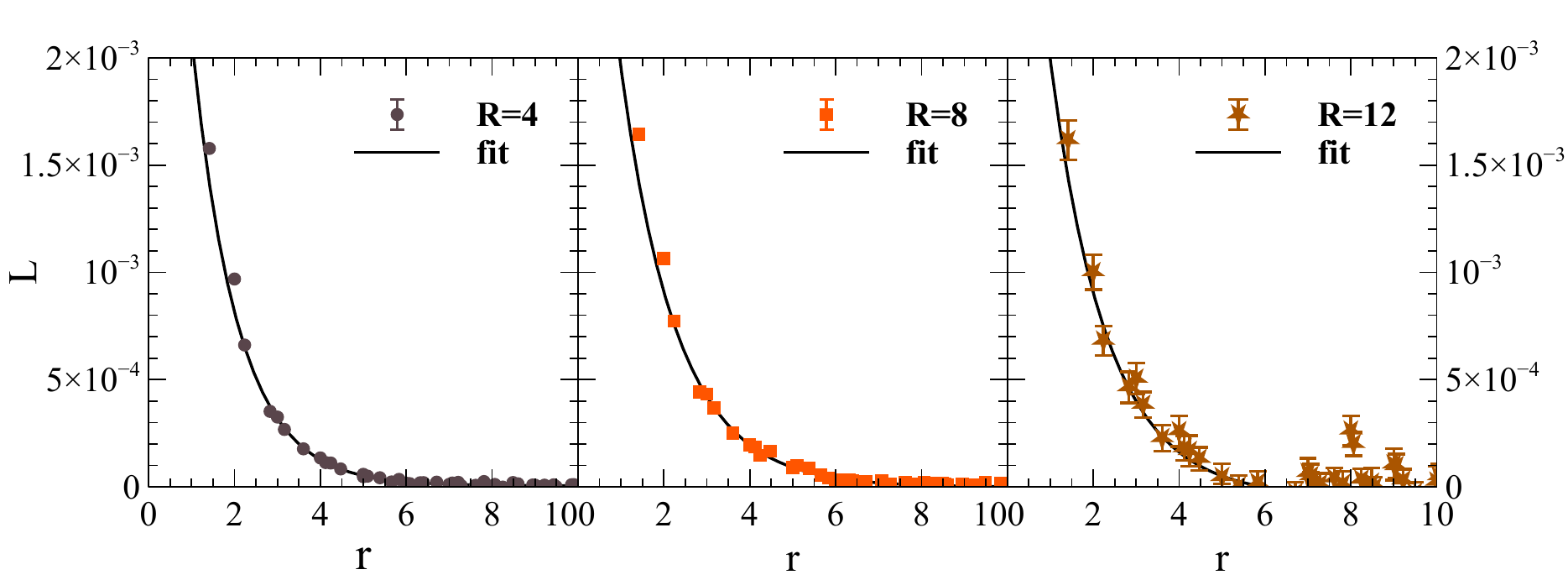}
\caption{Results and exponential fit to the large distance behavior in the charge planes.} \label{charge fit}
\end{figure}

Finally, as depicted in Fig. \ref{charge fit}, we depart from the mediator plane and study the charge planes. We fit the behavior of the Lagrangian --- at large distance from the charges --- to the exponential ansatz:
$\mathcal{L} \sim \mathcal{L}_0 e^{ - 2 \frac{r}{\lambda}}$. This result supports our hypothesis that, far from the charges, the field is screened with the same constant $\lambda$ parameter (similar to London Length) in the range $0.22\ \mbox{to}\ 0.24 fm$.

\section{Conclusion}
   We confirm earlier results \cite{Bakry:2010zt,Bakry:2010sp} that point to a  logarithm widening in the
   mediator plane of the flux-tube, in agreement with the results of quantum string models \cite{Luscher:1980iy}.
   Moreover, we also show that, at large quark-antiquark distances, the flux tube profile is not
   Gaussian but exponential as in a superconductor
   \cite{Cardoso:2013lla}. 
   This behavior is particularly prominent close to the charges.
   By studying this region, we obtain results that are consistent with a penetration length $\lambda$ of $2.2 - 2.4 fm$,
   corresponding to a dual gluon mass $\mu$ from $0.8$ to $0.9\,GeV$.

\begin{table}
	\centering
	\begin{tabular}{cccc}
  $R [a]$ &  $10^3 {\cal L}_0 $ & $\lambda[a]$ & $\chi^2/dof$\\ \hline
  4 &  5.3917  $\pm$ 17.468  & 2.1088 $\pm$ 0.1212 & 4.8315\\
  6 & 4.3832  $\pm$ 20.748  & 2.4803 $\pm$ 0.1376 & 2.1892\\
  8  & 4.2056  $\pm$ 11.041  & 2.6118 $\pm$ 0.1788 & 0.9665\\
  12& 5.6257  $\pm$ 36.337  & 2.2695 $\pm$ 0.5437 & 2.5743
	\end{tabular}
\caption{Calculated squared widths, obtained from the fit results. \label{sqwidths}}
\end{table}

  \section{Acknowledgments}
  We thank Martin L\"{u}scher,  Uwe-Jens Wiese, and Pedro Sacramento for enlightening discussions on flux tubes.
  This work was supported by Portuguese national funds through
  FCT - Funda\c{c}\~{a}o para a Ci\^{e}ncia e Tecnologia, projects PEst-OE/FIS/UI0777/2011,
  CERN/FP/116383/2010 and CERN/FP/123612/2011. 

  Nuno Cardoso is also supported by FCT under the contract SFRH/BD/44416/2008 and 
  Marco Cardoso under the contract SFRH/BPD/73140/2010.

We also acknowledge NVIDIA support with an Academic Partnership Program and a CUDA Teaching Center Program.

\end{document}